\documentclass[10pt,twocolumn]{article}
\usepackage[top=2cm, bottom=2cm, left=2cm, right=2cm]{geometry}
\usepackage{times}  %
\usepackage[sort&compress,numbers]{natbib}  %
\usepackage[hyphens]{url}
\usepackage{flushend}
\usepackage[hyphens]{url}  %
\usepackage{graphicx} %
\usepackage{tikzsymbols}
\usepackage{graphicx} 
\usepackage{xcolor}
\usepackage{xspace}
\usepackage{xurl}
\usepackage{wrapfig}
\usepackage{booktabs}
\usepackage{comment}
\usepackage{subfigure}
\usepackage{multirow}
\usepackage{amsmath,amsfonts}
\usepackage{bm}
\usepackage{tabularx} 

\usepackage{sectsty}
\sectionfont{\bfseries\Large\raggedright}

\let\oldbibliography\thebibliography
\renewcommand{\thebibliography}[1]{%
  \oldbibliography{#1}%
  \setlength{\itemsep}{2pt}%
}

\usepackage[hang,flushmargin]{footmisc}

\usepackage[compact]{titlesec}
\titlespacing*{\section}{0pt}{*3}{3pt}
\titlespacing*{\subsection}{0pt}{*2}{2pt}
\usepackage{xspace}

\usepackage[skipabove=5pt,skipbelow=5pt,roundcorner=10pt,framemethod=TikZ]{mdframed}

\makeatletter
\def\url@leostyle{%
  \@ifundefined{selectfont}{\def\UrlFont{}}%
  {\def\UrlFont{}}%
}
\makeatother
\usepackage[hyphenbreaks]{breakurl}

\definecolor{darkgreen}{RGB}{0, 100, 0}
\definecolor{linkcol}{rgb}{0.3,0,0}
\definecolor{citecol}{rgb}{0.3,0,0}
\definecolor{urlcol}{rgb}{0.3,0,0}

\usepackage[bookmarks=true, bookmarksnumbered=true, colorlinks=true, linkcolor=linkcol, citecolor=citecol, urlcolor=urlcol, hypertexnames=true]{hyperref}

\newcommand{\descr}[1]{\smallskip\noindent\textbf{#1}}
\definecolor{DarkGreen}{RGB}{0,100,0} 
\definecolor{Darkblue}{RGB}{0, 0, 160} 

\newif\ifcomment
\commenttrue
\ifcomment
\newcommand{\hs}[1]{{\bf \textcolor{brown}{HS: #1}}}
\newcommand{\hh}[1]{{\bf \textcolor{red}{HH: #1}}}
\else
\newcommand{\hs}[1]{}
\newcommand{\hh}[1]{}
\fi

\newcommand{\name}{{\textsc{Sentinel}}\xspace}

\makeatletter
\def\url@leostyle{%
  \@ifundefined{selectfont}{\def\UrlFont{\small}}%
  {\def\UrlFont{}}%
}
\makeatother
\begin{document}

\sloppy
\title{\bf \name: A Multi-Modal Early Detection Framework for Emerging Cyber Threats using Telegram}

	\author {
	Mohammad Hammas Saeed\textsuperscript{\rm 1}, and
	Howie Huang\textsuperscript{\rm 1} \\[1ex]
%
	\textsuperscript{\rm 1}George Washington University
}

\date{}

\maketitle

\begin{abstract}
Cyberattacks pose a serious threat to modern sociotechnical systems, often resulting in severe technical and societal consequences.
Attackers commonly target systems and infrastructure through methods such as malware, ransomware, or other forms of technical exploitation.
Most traditional mechanisms to counter these threats rely on post-hoc detection and mitigation strategies, responding to cyber incidents only after they occur rather than preventing them proactively.
Recent trends reveal social media discussions can serve as reliable indicators for detecting such threats.
Malicious actors often exploit online platforms to distribute attack tools, share attack knowledge and coordinate.
Experts too, often predict ongoing attacks and discuss potential breaches in online spaces.

In this work, we present \name, a framework that leverages social media signals for early detection of cyber attacks.
\name aligns cybersecurity discussions to real-world cyber attacks leveraging multi modal signals, i.e., combining language modeling through large language models and coordination markers through graph neural networks.
We use data from 16 public channels on Telegram related to cybersecurity and open source intelligence (OSINT) that span 365k messages.
We highlight that social media discussions involve active dialogue around cyber threats and leverage \name to align the signals to real-world threats with an F1 of 0.89.
Our work highlights the importance of leveraging language and network signals in predicting online threats.

\end{abstract}

\section{Introduction}
Modern cyber attacks are increasingly sophisticated, often orchestrated by distributed and covert actors who operate across loosely connected online ecosystems~\cite{appiah2020organizational,agrafiotis2018taxonomy,mezzour2014global}.
A cyber attack is a deliberate attempt by individuals or groups to gain unauthorized access to computer systems, networks, or devices with the intent to steal data, disrupt operations, or cause damage.
These attacks can take many forms (e.g, phishing scams or denial-of-service attacks) and pose serious risks by disrupting critical operations and often compromising sensitive data.
As a result, organizations often face shutdowns and long-term business setbacks, which includes financial losses~\cite{kamiya2018impact,tariq2018impact,chithaluru2020cyber}.
Beyond economic consequences, cyber attacks also threaten national security by targeting government and defense systems.
For example, in 2020, cyber criminals took over Twitter accounts of influential people (e.g., Barack Obama, Kim Kardashian West, Jeff Bezos, and Elon Musk) by using scamming techniques such as impersonating Twitter's Information Technology department and stole over \$118,000 worth of bitcoin~\cite{twitter2020report}.
These threats are becoming even more concerning as Artificial Intelligence (AI) is now being used as an additional tool in conducting and leading cyber attacks~\cite{guembe2022emerging,anthropicDisruptingFirst}.

\descr{Motivation.}
Social media platforms increasingly emerge as critical repositories of actionable intelligence for anticipating and mitigating cyber attacks.
A large body of work has shown that social media signals can serve as reliable early indicators of real-world cyber threats~\cite{alketbi2024cyber, abbes2025early, altalhi2021survey, sapienza2017early, marinho2023cyber, mardassa2024cyber, shu2018understanding}.
Shifts in language, sentiment and narratives often precede attacks, where not only attackers leverage social media to plan attacks but also experts discuss possibility of attacks and strategies.
These signals, however, are scattered across noisy and diverse social media discussions.
Unlike traditional cybersecurity analysis, which is often constrained to technical indicators and post-incident reporting, social media offers real-time, user-generated data that reflects both malicious activity and community-driven awareness.
This positions social media platforms as valuable complements to conventional threat intelligence sources.
In addition, the networked and temporal dynamics of social media make it suitable for predictive modeling.
For example, sudden increases in exploit-related discussions, coordinated reposting of attack toolkits, or semantic shifts in language usage can all signal the growing popularity of a vulnerability.
Individual users also frequently report anomalous behaviors in real time (e.g., phishing attempts or malware infections).
While such reports are often anecdotal and noisy, their aggregation across large-scale datasets can enable the identification of macro-level patterns.
This intelligence aids traditional detection pipelines by giving situational awareness to a global set of observations.
Additionally, these signals are inherently multimodal, i.e., emerge from patterns across communities, language and social signals which makes it important to consider multiple modalities when developing threat prevention systems.

\descr{Our Approach.}
Prior research leveraging social media signals for cyber attack prediction has demonstrated promising results~\cite{alketbi2024cyber, zhao2020timiner, marin2018mining}.
However, with increasingly adaptive attacker strategies, the need for early and proactive prediction has become more critical, especially since most defense responses are post-hoc.
Our work aims to bridge this gap by leveraging multimodal signals derived from social media to enable timely detection and mitigation of exploitative behaviors.

Past works have used several NLP techniques (e.g., BERT-based models) on social media feeds such as Twitter, hacking forums and dark web to map potential threats.
Popular language-based model techniques include Term Frequency-Inverse Document Frequency (TF-IDF) based representation (e.g., \cite{altalhi2021survey, alketbi2024cyber, marinho2023cyber}), Bidirectional Encoder Representations (BERT)~\cite{abbes2025early}, word embeddings (e.g., word2vec)~\cite{zhao2020timiner} or dictionary-based word frequency counts~\cite{almahmoud2023holistic}.
Similarly, sentiment or stance analysis have also been used as isolated predictors~\cite{shu2018understanding, marinho2023cyber, sapienza2017early}, while some attention has been given to network markers~\cite{sarkar2019predicting}.

In this work, we present \name, an early forecasting framework for cyber threats using both language and network features.
\name first encodes daily aggregated online discussions through semantic embeddings and then constructs a temporal-semantic graph of days to capture structural dependencies.
Next, \name applies GraphSAGE to generate graph embeddings, which along with the text-based embeddings are fed into a classifier for predicting cyber events on a given day.
Our results highlight that using structural features in combination with language features can improve predictive performance.
Therefore, we show that classical models with a unimodal focus (e.g., language features in isolation) can be significantly improved by incorporating additional modalities.

\descr{Research Questions} 
Overall, we aim to answer the following research questions through our work:

\begin{itemize}
    \item \textbf{RQ1:} Do we find indicators and discussion of cyber attacks on Telegram?
    \item \textbf{RQ2:} Do we observe a change in the language of cyber security communities over time?
    \item \textbf{RQ3:} Can we leverage the multimodal signals (i.e., language markers and network characteristics) to develop a predictive model for real-world cyber attacks?
\end{itemize}

\descr{Ethics Statement.}
All data analyzed were collected from publicly accessible channels or groups that do not require authentication or membership.
We did not collect, access or share any private messages or sensitive personal data.
We also anonymized usernames and profile metadata prior to our analysis.

\descr{Contributions.}
Through this work, we make several key contributions for cyber threat prediction.
Building upon our research questions, we summarize our contributions as follows:

\begin{enumerate}
    \item We collect a first-of-its-kind dataset of cyber-focused discussions on Telegram that can be leveraged for cybersecurity analyses.
    \item We identify that social media discussions contain signals around cyber threats and there are observable changes in activity as it relates to real cyber-incidents.
    Overall, we analyze 365k messages from Telegram across 16 groups.
    We also find that language in these communities evolves over time.
    \item We present \name, a hybrid model that uses network and semantic features to predict cyber attacks. Through our analysis, we highlight the importance of multimodal features, achieving an F1-score of 0.89 and accuracy of 0.91 in our best implementation.
    \name encodes the messages into temporally aligned daily semantic embeddings using OpenAI text-embedding model.
    These embeddings are aggregated across the groups and paired with real-world cyber event data to enable prediction.

\end{enumerate}

\descr{Paper Organization.}
The rest of the paper is organized as follows.
The next section describes our dataset.
In Section 3, we introduce the several components and design of our system \name.
Next, in Section 4, we present our analysis of Telegram data and the evaluation of \name for predicting real cyber incidents through social media signals.
In Section 5, we contrast other works related to our research followed by a discussion of the implications of our results and highlight the importance of our multimodal design in Section 6.
Lastly, we conclude our paper in Section 7 offering a promising path for developing early-warning systems for cyber threats.

\begin{table}[t]
\centering
\begin{tabular}{|l|c|c|c|}
\hline
\textbf{Group} & \textbf{Active Period} & \textbf{Messages} \\
\hline
cybersecurityexperts & March 2019 -- June 2025 & 233,226 \\ \hline
itsectalk & Jan 2017 -- April 2025 & 46,386 \\ \hline
cyber\_security\_feed & Aug 2020 -- June 2025 & 27,905 \\ \hline
BugCrowd & May 2020 -- May 2025 & 19,962 \\ \hline
WokeIntelDrops & July 2020 -- March 2023 & 13,152 \\ \hline
PHOfficial & Feb 2018 -- June 2025 & 7,732 \\ \hline
cissp & Aug 2017 -- June 2025 & 7,492 \\ \hline
cybdetective & Aug 2021 -- June 2025 & 3,019 \\ \hline
cloudandcybersecurity & June 2020 -- June 2025 & 2,151 \\ \hline
hackers\_asylum & Jan 2023 -- June 2025 & 1,750 \\ \hline
hackersworldunited & May 2022 -- May 2025 & 543 \\ \hline
HackingBlogsGroup & April 2024 -- May 2025 & 791 \\ \hline
joinhackingarmy & July 2023 -- May 2025 & 716 \\ \hline
bellingcat\_en & Oct 2018 -- May 2025 & 435 \\ \hline
itsecalert & Jan 2016 -- Dec 2023 & 125 \\ \hline
espyOSINT & March 2022 -- June 2022 & 86 \\ \hline
\hline
\textbf{Total} &  & \textbf{365,471} \\
\hline
\end{tabular}
\caption{Dataset breakdown from different groups}
\label{tbl:dataset}
\end{table}

\begin{table}[h!]
\centering
\begin{tabular}{|l|c|}
\hline
\textbf{Attack Type} & \textbf{Count} \\
\hline
malware & 2,085 \\ \hline
other & 1,496 \\ \hline
vulnerability & 931 \\ \hline
account takeover & 641 \\ \hline
targeted attack & 601 \\ \hline
ransomware & 443 \\ \hline
ddos & 229 \\ \hline
scam & 141 \\ \hline
coordinated inauthentic behavior & 139 \\ \hline
misconfiguration & 57 \\ \hline
malicious script injection & 50 \\ \hline
business email compromise & 28 \\ \hline
malvertising & 28 \\ \hline
defacement & 18 \\ \hline
sqli & 13 \\ \hline
credential stuffing & 13 \\ \hline
brute-force & 11 \\ \hline
deepfake & 9 \\ \hline
dns hijacking & 8 \\ \hline
flash loan & 6 \\ \hline
crypto drainer & 6 \\ \hline
crypto scam & 5 \\ \hline
\hline
\end{tabular}
\caption{Cyberincidents data breakdown by attack type}
\label{tbl:cyber_incidents}
\end{table}

\begin{figure*}
    \centering
    \includegraphics[width=0.8\textwidth]{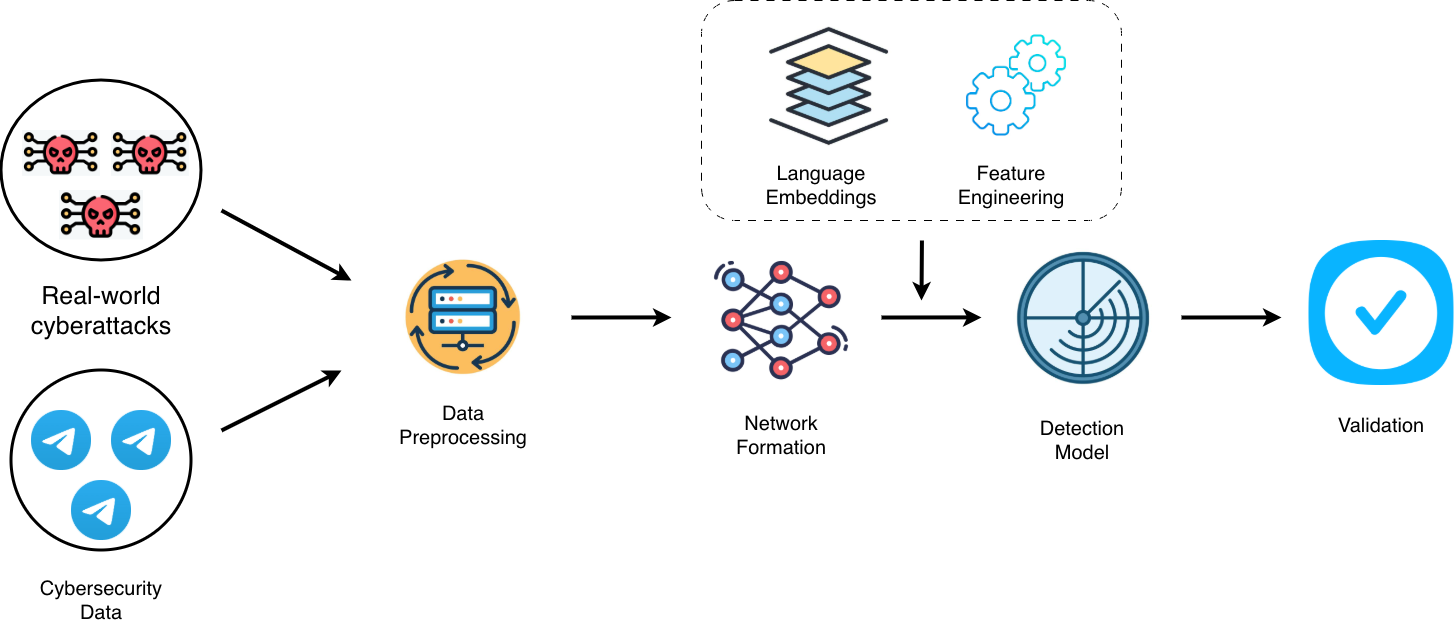}
    \caption{Overview of \name: The system is fed raw cyber-focused Telegram messages and real-world timeline of cyberincidents.
    Next, it cleans the data into daily documents to reflect evolving discourse.
    Each day is modeled in the graph as a node connected to its neighboring days, allowing information to flow over time.
    \name generates text embedding using a transformer model.
    GraphSAGE is then applied to produce contextual temporal embeddings for \name to encode how discussion evolves, escalate or decay over time.
    The hybrid representation from graph and text embeddings is then fed into a supervised classifier to detect cyber events.}
    \label{fig:system}
\end{figure*}

\section{Data}
For the purposes of this work, we use data from cybersecurity discussion channels on Telegram.

\descr{Telegram.}
Telegram is a cloud-based instant messaging application that offers communication with support for text messages, multimedia sharing, voice and video calls, large group chats, and public broadcast channels.
Table~\ref{tbl:dataset} gives a complete list of dataset we use in our study.
Unlike many messaging platforms, Telegram allows massive group sizes, up to 200,000 members, and supports file sharing of up to 2 GB per file.
One of its features is its cross-platform synchronization, enabling access to chats across phones, tablets, and desktops.
The platform has gained popularity for its openness, extensive bot support, and developer-friendly API, making it a hub for communities, activists, and businesses.
Telegram is also widely used across the world for personal communication, public information sharing, and even large-scale social movements~\cite{thomas2022comprehensive}.
It is also a valuable source for cyber threat intelligence because it is now being used ss gathering places for malicious actors~\cite{roy2024darkgram}.

\descr{Hackmageddon.}
We utilize Hackmageddon as our source of real-world cyber incidents, which has also been used by prior work on cyber threat prediction~\cite{dalton2017improving,hajizada2023gaps,werner2017time}.
Hackmageddon is a cybersecurity intelligence website that tracks cyber attacks and breaches worldwide from open source data.
It provides detailed timeline of cyber attacks along with their motivation (e.g., cyber attack, hacktivism) and sometimes target sector (such as government or finance) along with the attack technique that has been used.
The attack techniques include DDoS, phishing, malware and others.
The data on Hackmageddon is collected from publicly available reports and news sources which serves as an open-source intelligence (OSINT) repository for researchers.
Overall, we compile a ground truth dataset of 6,957 cyber events from Hackmageddon~\cite{passeri2025hackmageddon}.
The complete breakdown of the types of attacks is given in Table~\ref{tbl:cyber_incidents}.
Attacks that are marked ``Unknown'' by Hackmageddon or appear less than 5 times in the data are compiled in the ``other'' category.
We also combine the different mentions of ``CVE'' and ``vulnerabilities'' into a single descriptor, i.e., ``vulnerability.''

\begin{figure*}[t]
    \centering
    \begin{minipage}[t]{0.48\linewidth}
        \centering
        \includegraphics[width=\linewidth]{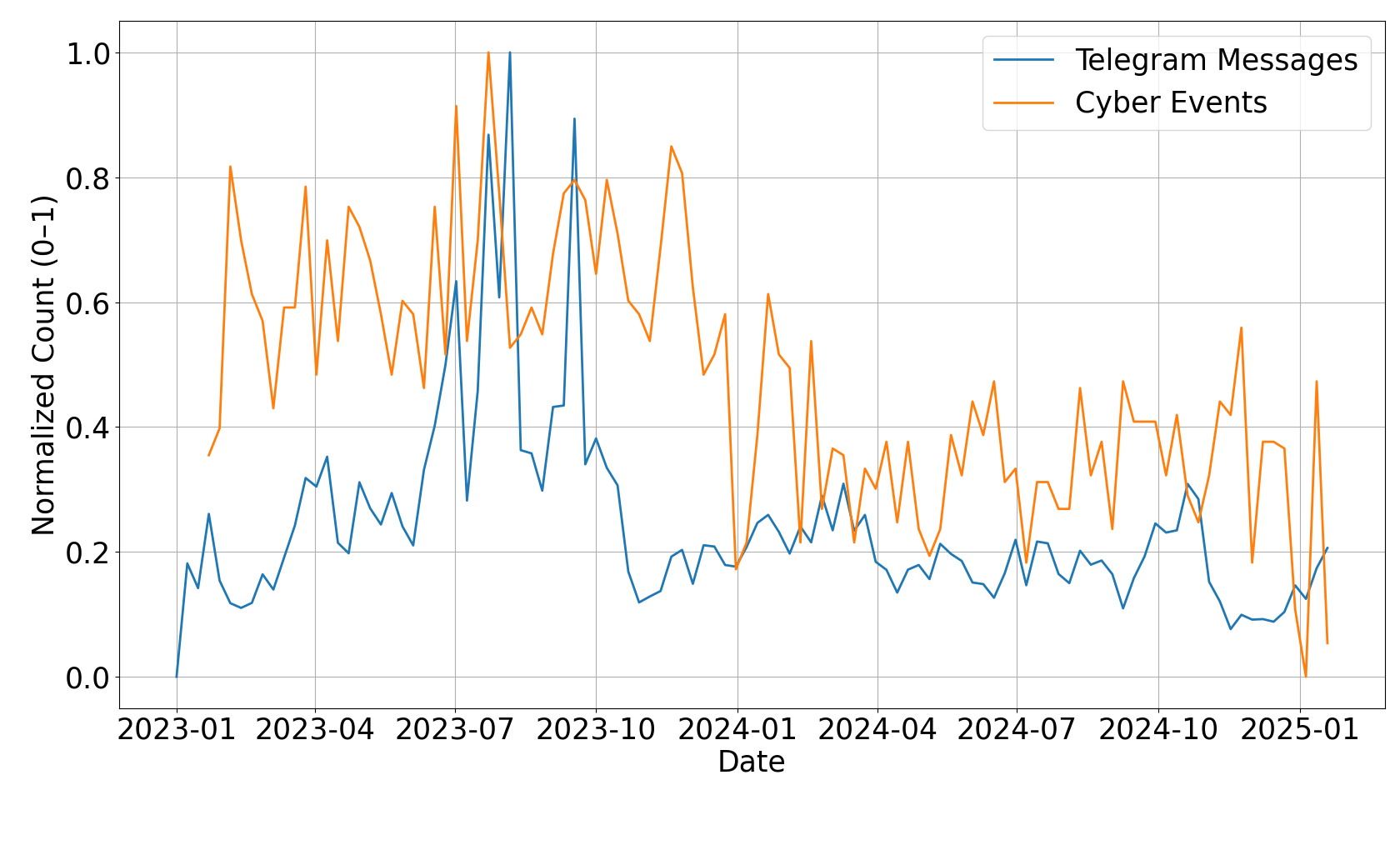}
        \caption{Weekly trends in \textcolor{blue}{Telegram messages} and reported \textcolor{orange}{cyberattack events}}
        \label{fig:time_alignment}
    \end{minipage}\hfill
    \begin{minipage}[t]{0.48\linewidth}
        \centering
        \includegraphics[width=\linewidth]{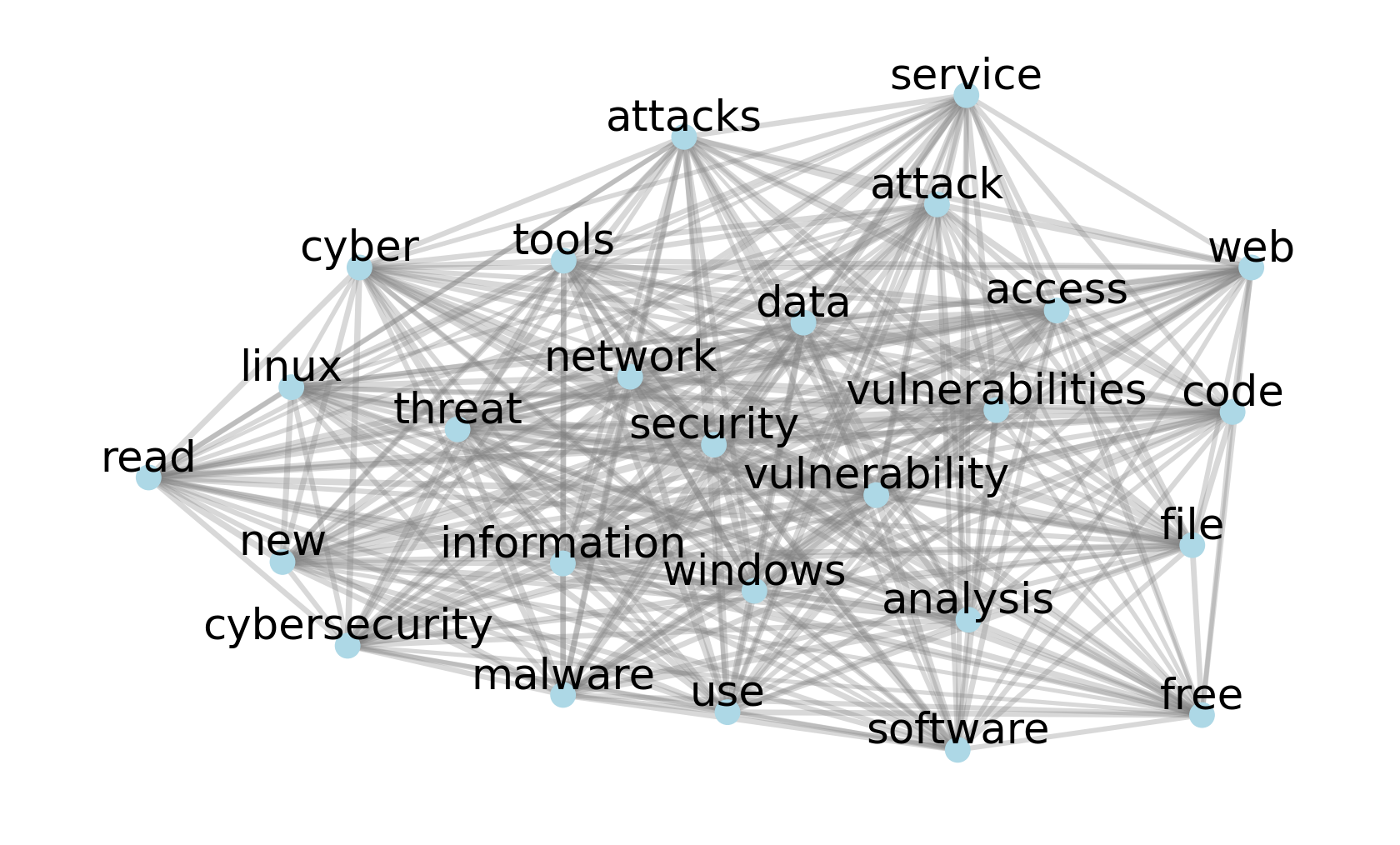}
        \caption{Keyword co-occurrence graph from group messages}
        \label{fig:keywords_graph}
    \end{minipage}
\end{figure*}

\section{Methodology}\label{sec:Method}
\name predicts the temporal relationship between online cyber discussions and real-world cyber incidents from open-source events.
Figure~\ref{fig:system} gives the overview of the system.
It operates through several interconnected stages, from data preprocessing and embedding generation to supervised learning and alignment.

\descr{Data Collection and Preprocessing.}
We first collect JSON-formatted Telegram group data.
To query the groups, we use the Telethon API\footnote{\url{https://docs.telethon.dev/en/stable/concepts/full-api.html}}.
The data is stored such that each file represents a distinct Telegram group or channel that discusses cybersecurity-related topics.
The messages contain fields such as date, message, and metadata.
To ensure a consistent input format, messages are obtained from the text field and missing entries are discarded.
Only messages posted after January 1, 2023 because Hackmageddon data starts from 2023 and it allows us to focus the analysis on recent trends.
Each message is associated with its posting date, forming a chronological dataset of text content over time.
Messages are grouped by date, producing a list of all messages for each day within each Telegram group.

\descr{Embedding Generation.}
To transform the textual content into a machine-understandable representation, \name employs the OpenAI \textit{text-embedding-3-small} model, a state-of-the-art transformer-based embedding generator.
The \textit{text-embedding-3-small} model achieves better performance on benchmark tasks than the earlier \textit{text-embedding-ada-002} model.
We also use the small version to minimize computation, memory and storage cost.
Each piece of text is converted into a 1536-dimensional vector that captures the contextual meaning of the message beyond simple word frequency.
The computed vectors capture the semantics of the text so systems can compare, search, cluster, or classify language efficiently.
We compute the embeddings in batches of 50 messages using the \textit{get\_embeddings\_batch()} function, which cleans the text and sends it to the OpenAI API for vectorization.

For each Telegram group and each day, all message embeddings are averaged to create a daily semantic representation of that groups discussion activity.
This daily embedding vector serves as a condensed snapshot of the group's overall topic and tone for that day.

\descr{Aggregation Across Groups.}
After all groups are processed, the individual daily embeddings are merged into a single unified matrix.
Each row represents a calendar day, and each column corresponds to the combined embedding features across all groups.
Missing days are filled with zero values to maintain temporal continuity.
This results in a daily time series of aggregated semantic signals.

\descr{Network Formation.}
Although text embeddings capture semantic information, they do not inherently model temporal and structural dependencies.
Cyber attacks often show temporal structure (e.g., evolving discussions and ongoing coordination) which precedes the incident.
To incorporate such dependencies, \name constructs a graph where each node represents daily semantic embeddings.

The graph \(G = (V,E)\) is defined such that each node \(v_t \in V\) corresponds to a day \(t\).
The feature vector associated with \(v_t\) is the text embeddings for that day.
Thus, each node inherits the semantic characteristics of cyber discussions occurring on that day.
We define edges based on relation between days.
First, we add forward edges each day to the next for capturing the sequential dependency that discussions of one day has on the following day.
Next, to model longer-range periodic behavior, i.e., weekly influence, we also add edges \(t \rightarrow t+7\).
These edges allow the graph neural network to aggregate information over longer windows and capture patterns that unfold over several days.

\name uses GraphSAGE~\cite{hamilton2017inductive} which is designed for inductive node representation making it well suited for tempporal data.
Unlike classical Graph Convolutional Networks (GCNs), GraphSAGE does not require to be retrained with the entire graph when new data is added making it flexible for evolving patterns in cyber discourse.
GraphSAGE computes node representations iteratively.
For each layer, node embeddings are updated by aggregating information from neighbors.
We employ a 2-layer model where each layer uses ReLU activation and the loss function is a weighted binary cross entropy.
After training, GraphSAGE produces contextualized node representations.
These embeddings involve temporal dynamics otherwise invisible to a purely text-based model.

\descr{Integration of Cyber Event Timelines.}
To link online activity with real-world cyber incidents, \name imports a series of cyber event timeline.
The timelines from Hackmageddon include event-level metadata such as attack type, target organization, and date of occurrence.
The model standardizes date formats and filters for valid records containing the Date Occurred field.
Only events dated 2023 and onward are included, ensuring temporal overlap with the Telegram message data.
The events are then resampled into a daily frequency.
The number of attacks per day is counted, where each value indicates the daily count of reported cyber incidents.

\descr{Classification and Evaluation.}
The model aims to aligns the embedding based signal (X) with the cyber incident timeline (y).
We perform binary classification such that, \texttt{1} indicates that a cyberevent took place that day and \texttt{0} indicates no recorded event.
\name learns to predict the likelihood of a cyber event given the semantic profile of the online conversations from that day.
For validation, we divide our data into a (70/30) split using stratified sampling to maintain class balance.
A Random Forest Classifier is used to predict presence or absence of cyber incidents, given the embedding signal.

\begin{figure}[t]
    \centering
        \includegraphics[width=\linewidth]{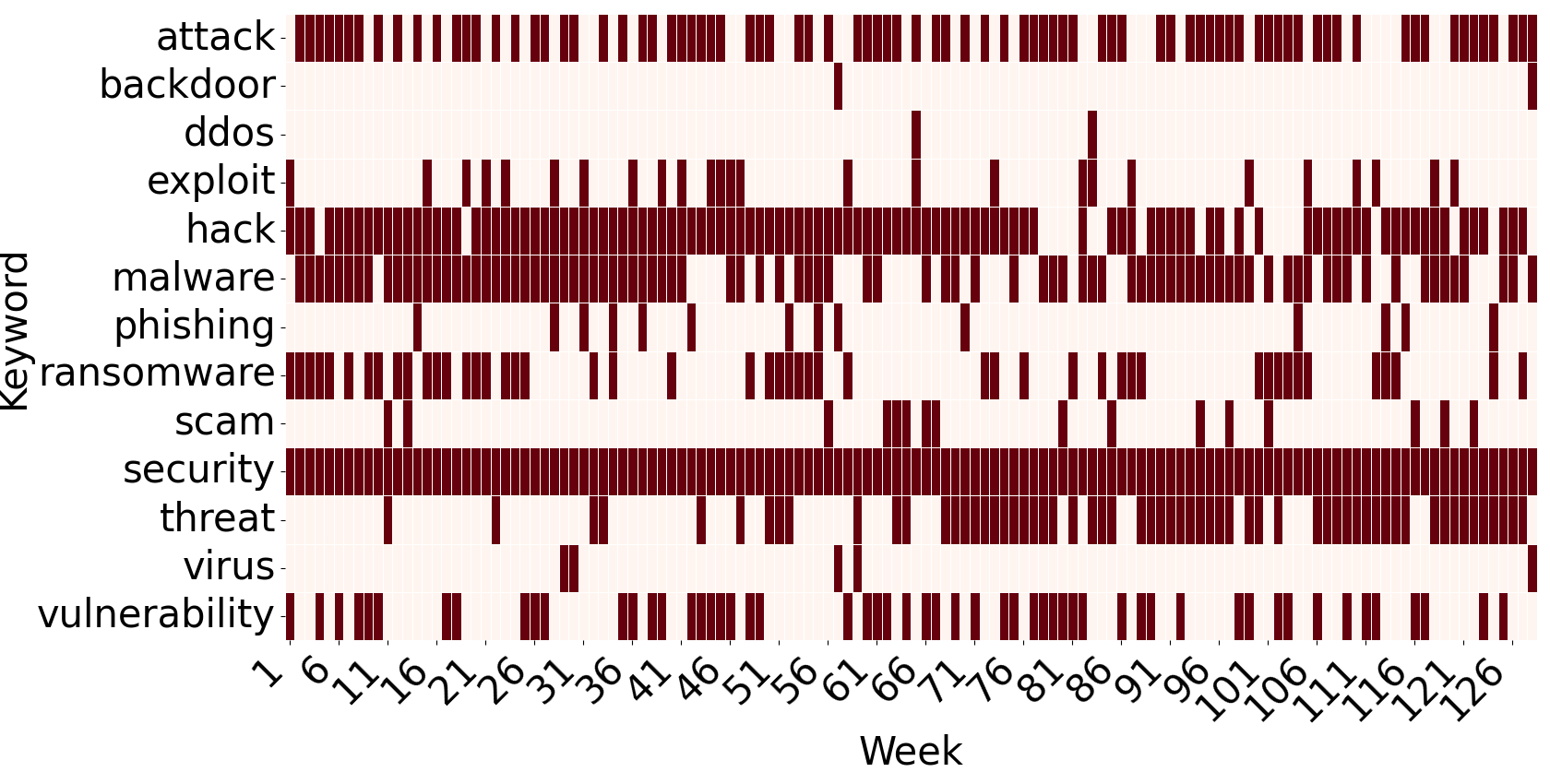}
        \caption{Trends of keywords appearing as top TF-IDF terms}
        \label{fig:keyword_heatmap}
\end{figure}

\begin{figure}[t]
\centering
\includegraphics[width=\linewidth]{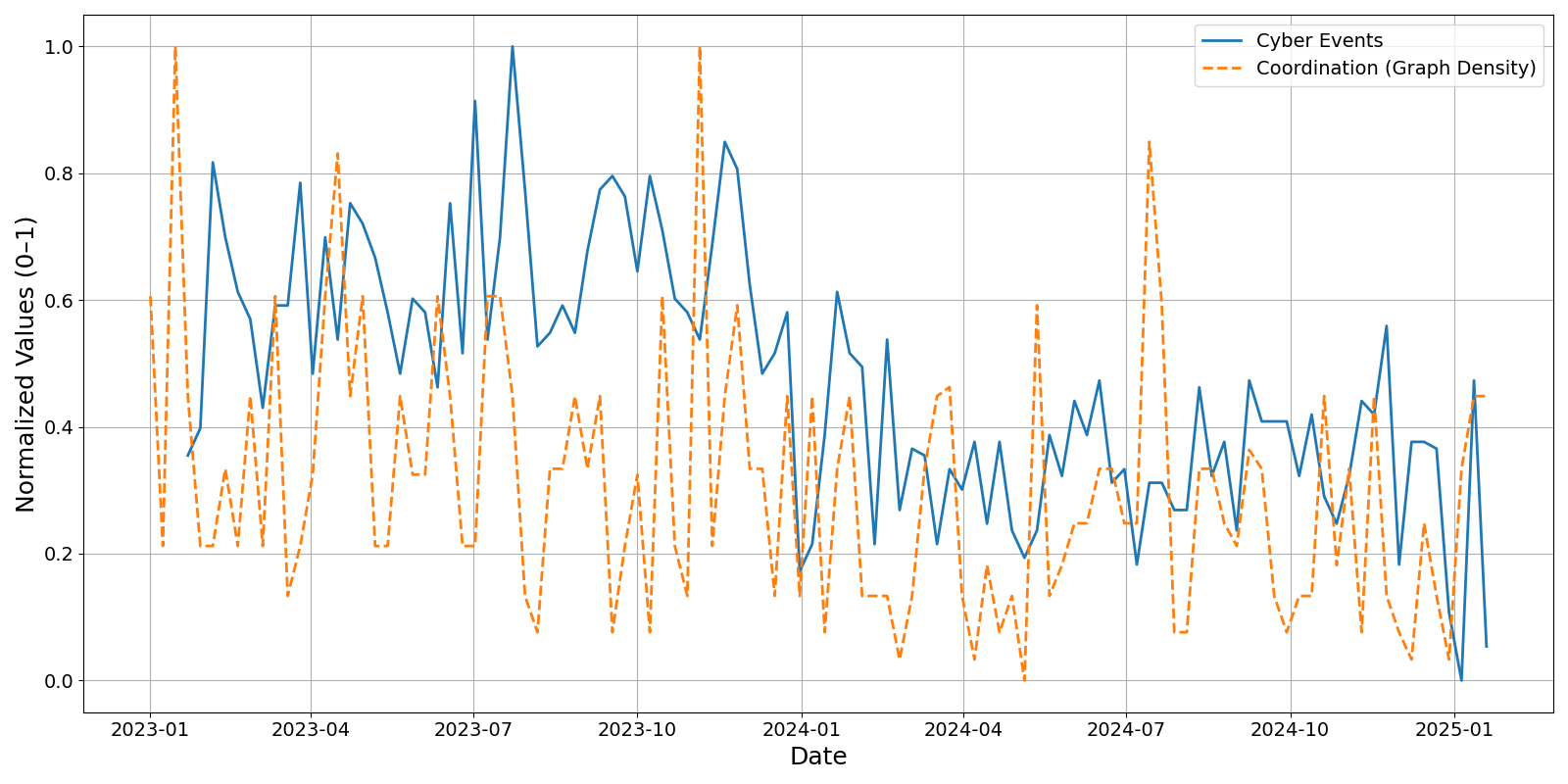}
  \caption{Graph density and cyber incidents over time}
\label{fig:graph_metrics}
\end{figure}

\section{Results}
In this section, we present our findings for the three research questions.
We first analyze the Telegram data for early threat indicators (RQ1), followed by understanding evolving langauge in groups (RQ2) and lastly, we use \name for multimodal alignment to cyberattack prediction (RQ3).
Through our analysis, we highlight language and structural signals together provide a more comprehensive design for timely detection of cyber threats.

\subsection{RQ1: Early indicators and discussion of socially engineered cyber attacks}
We start by assessing whether fluctuations in Telegram discussion activity corresponds to documented real-world cyber incidents.
To investigate this, we construct weekly time series for: a) the volume of messages across all cyber-focused Telegram channels and b) the number of reported cyberattacks extracted from Hackmageddon. 
This allows us to examine temporal patterns in message volume aligning with specific attack events.
As illustrated in Figure~\ref{fig:time_alignment}, the normalized time series shows spikes in Telegram discussions often co-occur with reported cyber incidents, suggesting that chatter within expert communities increases with impending or active attacks.

Next, we analyze messages within communities to identify the main topics discussed.
For this, we first filter messages that contain seed cybersecurity keywords, i.e., Advanced Persistent Threats (or ``APT'') and Common Vulnerabilities and Exposures (or ``CVE'') to focus on discussions related to active threats and vulnerabilities.
Then, we build a word co-occurrence matrix where each pair of words appearing in the same message increases their association count.
Using this matrix, we construct a graph where each node represents a unique word and edges represent co-occurrence strength.
The resulting network captures the contextual relationships between terms used in threat discussions.
Figure~\ref{fig:keywords_graph} shows that visualizing the top words in this network reveals frequent mentions of cybersecurity-relevant terms like ``attack,'' ``threat,'' ``vulnerability,'' ``tools,'' and ``access.'' 
The network shows active discussion around key themes in cyber attacks.
Another important point to note is that discussion around vulnerabilities is strongly connected to keywords like ``analysis'' and ``tools'' highlighting that discussions around attacks are nuanced and in-depth.

Next, we analyze important keywords in the data over time.
We compute the Term Frequency - Inverse Document Frequency (TF-IDF) of each keyword.
Figure~\ref{fig:keyword_heatmap} shows the keywords heatmap, which is a visual representation that tracks how prominent a given cyber threat-related keyword extracted from weekly messages.

The darker red hues indicate higher TF-IDF values, signifying that the term appeared more frequently and with greater importance relative to other words in that period.
This heatmap highlights spikes in certain keywords over time and how keywords gain or lose prominence over time making it important to understand evolution of language in these spaces.
We focus on threat related keywords such as ``malware,'' ``vulnerability'' and ``ddos'' amongst others.

To analyze group coordination, we also construct weekly graphs where nodes represent Telegram groups and edges connect groups that posted identical messages in the same week, using graph density as a quantitative measure of coordination.
This coordination metric, along with the aggregated weekly message volume and cyber event counts, are normalized and plotted.
As shown in Figure~\ref{fig:graph_metrics}, we find consistencies in the initial time series revealing likely temporal relationships between activity and group posting behavior.

\begin{figure}[t]
    \centering
        \includegraphics[width=\linewidth]{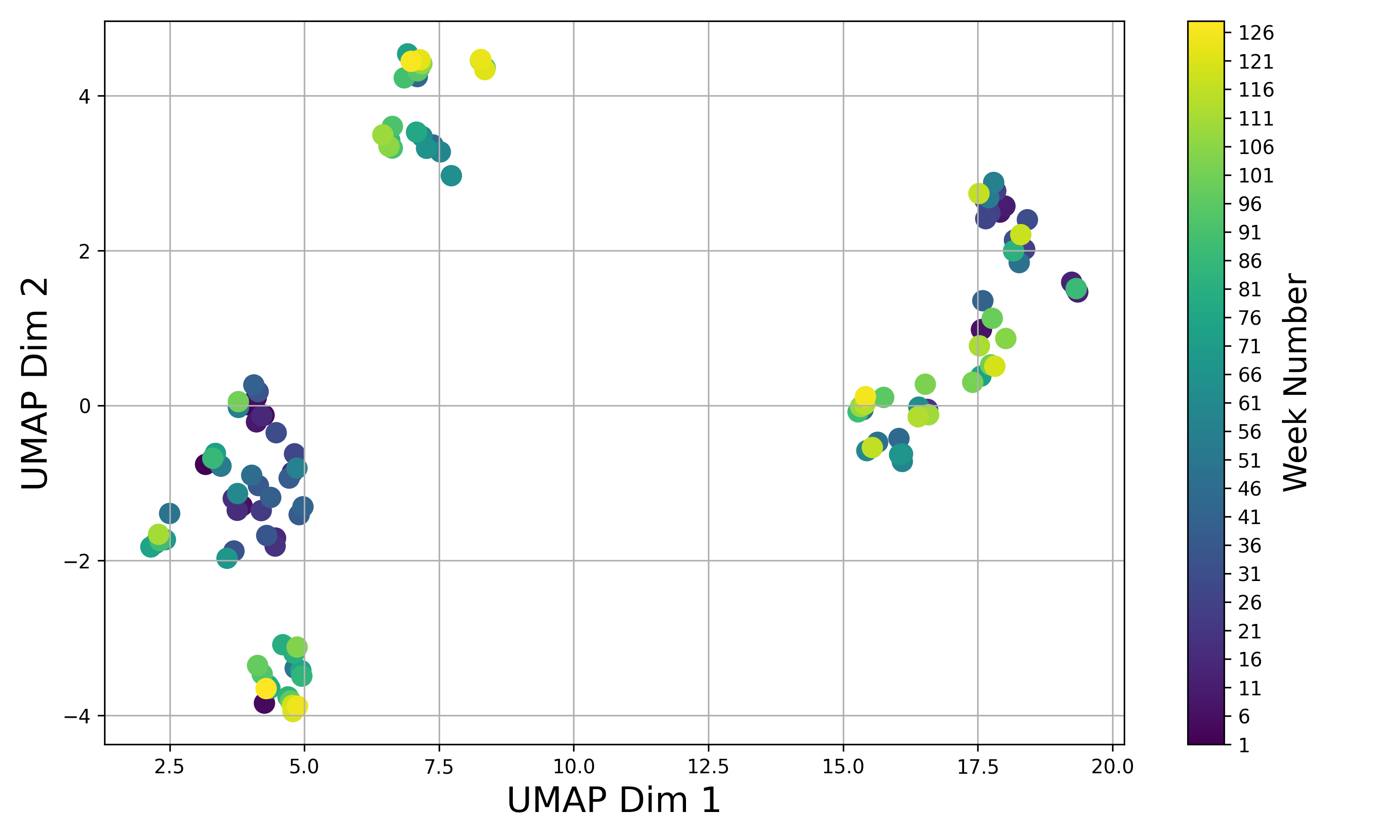}
        \caption{2D UMAP showing concept drift in semantic embeddings}
        \label{fig:umap}
\end{figure}

\subsection{RQ2: Evolving language in social media discussions}
To understand how language shifts over time, we analyze the semantic evolution of weekly data.
Figure~\ref{fig:umap} shows the Uniform Manifold Approximation and Projection (UMAP) over multiple weeks based on Telegram data.
We first average high-dimensional sentence embeddings of the weekly attack keywords, condensing the semantic content of each week's vocabulary into a single representative vector.
Next, UMAP then projects these averaged embeddings into a two-dimensional space designed to preserve semantic similarities, so that weeks with closely related themes trend near each other while semantically distinct weeks appear farther apart.
This visualization maps the high-dimensional semantic space of weekly discussions into a 2-dimensional representation.
This dimensionality reduction enables us to observe how the overall ``meaning'' of discussions changes over time and to identify periods where discourse shifts abruptly or stabilizes around particular themes.
We find the discussions around certain keywords to cluster and disperse over time showing that the nature of threat discourse evolves over time and also is contextually similar in certain time periods.

\subsection{RQ3: Aligning multimodal signals to timelines of real-world cyber attacks}
Lastly, we use \name to align signals with real-world attacks.
\name processes the conversations within Telegram groups.
As outlined in Section~\ref{sec:Method}, the Telegram message data is sourced from JSON files where each file corresponds to a distinct group.
Each message record contains a timestamp and textual content which is extracted.
\name then clusters messages by day within each group, creating daily message collections that capture the thematic focus of each group over time.
To transform raw text into numerical features, \name uses OpenAI's embedding model \textit{text-embedding-3-small}.
Messages are processed in batches to respect API constraints, with embeddings computed for each message.
For each group's daily message set, embeddings are averaged element-wise to create a representative semantic vector, summarizing the group's discourse for that day.

The daily group embeddings are concatenated across all groups to create a unified feature vector for each date which contains the combined semantic information available in Telegram discussions.
The embeddings are then used for classification.
A Random Forest Classifier is used to predict presence or absence of cyber incidents, given the embedding signal.
\name\textsc{-text} model achieves an F1 score 0.84 using only the semantic features.
Next, the \name\textsc{-hybrid} model includes the temporal graph of days to capture structural and periodic influence in addition to text embeddings which outperforms the pure language model.
We also compare \name with classical baseline models, i.e., a TF-IDF based model (TF) over the Telegram text and a Sentence-BERT (SBERT) model~\cite{reimers2019sentence} which is a transformer-based sentence embedding model fine-tuned for semantic similarity.
As shown in Table~\ref{tab:model_performance}, we find \name's multimodal design outperforming the text-based unimodal approaches, achieving an overall accuracy of 0.91 and an F1-score of 0.89.

\begin{table}[t]
\centering
\begin{tabular}{|l|c|c|c|c|}
\hline
\textbf{Model} & \textbf{Precision} & \textbf{Recall} & \textbf{F1} & \textbf{Accuracy} \\
\hline
TF & 0.72 & 0.85 & 0.79 & 0.84 \\
SBERT & 0.80 & 0.85 & 0.83 & 0.84 \\
\name\textsc{-text} & 0.83 & 0.85 & 0.84 & 0.85 \\
\name\textsc{-hybrid} & \textbf{0.90} & \textbf{0.89} & \textbf{0.89} & \textbf{0.91} \\

\hline
\end{tabular}
\caption{Alignment Performance}
\label{tab:model_performance}
\end{table}

\section{Related Works}
Recent research at the intersection of cybersecurity and machine learning has explored a wide range of techniques for anticipating or detecting cyberattacks by leveraging data from social media platforms, hacker forums, and dark web sources.
These works highlight the growing relevance of social media signals as predictors of cyber threats.

Several studies have mined hacker forums to extract features relevant for detecting or predicting cyber threats.
Mardassa et al.~\cite{mardassa2024cyber} conducted sentiment analysis on hacker forum posts using deep learning methods such as LSTM and GloVe embeddings.
Their study showed promise in classifying posts into positive and negative sentiment to anticipate cyber threats.
Similarly, Alketbi et al.~\cite{alketbi2024cyber} benchmarked traditional ML models (SVM, LR, RF, XGBoost) and deep models (LSTM, FNN) on hacker forum data labeled as hacking-related or not. Using TF-IDF, Word2Vec, and GloVe for feature extraction, the study underscored the diversity of hacker communities and the challenge of generalized threat detection.
Building upon this, Marin et al.~\cite{marin2018mining} identify ``key hackers'' in dark web forums using features like jargon frequency, content topic modeling, and user seniority, highlighting the social hierarchies that structure hacker communities.
In the same vein, more recent work~\cite{purba2025towards} has utilized GPT-based methods to extract cyber intelligence according to MITRE ATT\&CK framework from social media data.

Other works have used Twitter and similar platforms to predict or detect cyber threats.
For example, Khandpur et al.~\cite{khandpur2017crowdsourcing} introduce a crowdsourcing approach, mining social signals from Twitter (e.g., incident mentions) and comparing with Gold Standard Reports (e.g., Hackmageddon) for validation.
Shu et al.~\cite{shu2018understanding} use sentiment trends on Twitter grouped by categories like DDoS, phishing, and CVEs to understand threat behavior over time, while Marinho et al.~\cite{marinho2023cyber} create a classification system that maps Twitter content to MITRE ATT\&CK tactics using TF-IDF, NER, and co-occurrence of cybersecurity-specific terms.
Similarly, Altalhi et al.~\cite{altalhi2021survey} provided a survey of real-time Twitter-based cyberattack detection models, noting TF-IDF’s superior performance in early threat signal extraction.

To combine news, social media, and dark web for threat Intelligence, works like Zhao et al.~\cite{zhao2020timiner}(TIMiner) focus on harmonizing hacker forums, security bulletins, and news sources to extract structured threat descriptions using CNNs and Word2Vec.
Similarly, Goyal et al.~\cite{goyal1806discovering} explore ARIMA and Phased LSTM models using multi-source time-series data from dark web, blogs, and Twitter to predict cyber incidents at two organizations.
Sapienza et al.~\cite{sapienza2017early} built a four-stage filtering pipeline based on keywords using dark web and Twitter discussions to detect novel threat indicators.
Rahman et al.~\cite{rahman2023attackers} and Basheer et al.~\cite{basheer2021threats} offer surveys and taxonomies for automating threat intelligence extraction, focusing on text-based dark web sources.

The shift towards newer platforms such as Telegram and Reddit has also become visible in recent work.
Roy et al.~\cite{roy2024darkgram} analyze Telegram activity to trace cybercriminal coordination patterns, while Kuhn et al.~\cite{kuhn2024navigating} compared dark web posts with tweets for signal quality.
Vu et al.~\cite{vu2025yet} examine cyberattacks linked to geopolitical conflict by analyzing hacker forum posts and Telegram groups, using non-parametric Kruskal-Wallis and Dunn’s tests for inference.
Other works, such as the on by Charmanas et al.~\cite{charmanas2024content}, assesses user concerns and trends across information security threads, validating social media platforms such as Reddit to be legitimate cyber intelligence source.

Other works use event-centric models to forecast cyberattacks using incident timelines and structured logs.
Kannan et al.~\cite{kannan2024prediction} and Abu Bakar et al.~\cite{bakar2024ftg} focus on deep learning architectures (LSTM-RNN, FTG-Net-E) to detect anomalies in web traffic (CICIDS and UNSWNB datasets).
Ahmed et al.~\cite{ahmed2024rapid} and Goyal et al.~\cite{goyal1806discovering} apply time series models like linear regression and ARIMA to attack categories such as DDoS, botnet, and infiltration.
Almahmoud et al.~\cite{almahmoud2023holistic} combines Hackmageddon incident records with scientific literature and B-LSTM models to anticipate attack trends across news and social media and Abbes et al.~\cite{abbes2025early} uses BERT to classify Twitter activity before and during confirmed attacks.

\descr{Remarks.} 
While existing works have utilized a variety of data sources and traditional modeling techniques, \name uses a hybrid approach that combines large language models with a graph-based representation that allows for deeper understanding of unstructured text while simultaneously modeling the evolving network of users, messages, and topics.

\section{Discussion}
Our work demonstrates that combining semantic representation with temporal graph learning substantially improves the ability to characterize cyber threats from social media data.
We argue that language-based models can capture signals of discussion around security topics, however combining it with graph-based modeling allows us to captures how those discussions evolve over time.
Thus, a multimodal approach produces a more informative representation than either component in exclusivity.

\descr{Positive Implications.}
Our approach offers several advantages for threat detection and proactive defense.
One of the most important benefits is the ability to detect emerging threats earlier.
By leveraging text-based signals in addition to network evolution, \name places emphasis on detecting changes before major cyber events.
Proper use of the the model can help detect early-warning indicators of new vulnerabilities, exploit techniques, or coordinated attack activity.
Our goal is to strengthen defense measures and shift strategies from reactive to proactive.

\name also enhances situational awareness.
Large bodies of unstructured text, such as Telegram channels, OSINT feeds, exploit discussions, or threat reports are difficult for humans to digest in real time.
By converting this content into signals, the model provides a clearer picture of how the threat landscape is evolving.
Therefore, instead of relying solely on human intuition, decisions can be made based on measurable signals which could be otherwise hidden in large volumes of noise.
We also present a case for exploring multiple dimensions when dealing with social media data.
The importance of temporal and structural trends along with language modeling presents a stronger defense against cyber attackers.

\descr{Potential Risks.} 
While our approach offers great performance gains, there are several associated risks with such an approach.
Firstly, embeddings based models are susceptible to overinterpreting linguistic cues that can be correlated to hype or heightened discussion rather than pure operations-based discussion, which is why \name's design focuses on a hybrid representation to not rely on just one signal.
Secondly, temporal modeling can also lead to spurious associations, i.e., if discussions spike due to non-operational discussions (e.g., sensational news or viral content), \name could infer false sense of escalation.
Third, since cyber ecosystems are dynamic and adversarial in nature, threat actors may adapt their communication strategies in response, potentially inducing model drift or evasion.
Thus, training models on on open-source communications (i.e., Telegram discussions in our case) presents the possibility of biased data stream and potential blind spots in forecasting.

\descr{Safe Deployment.}
The safe deployment of \name requires several technical safeguards.
\name's predictions should be treated as decision support tools and not a deterministic indicator.
We envision human expert oversight in all automated outputs.
Secondly, all monitoring must comply with platform terms of service and data protection regulations with data being anonymized and unnecessary metadata being stripped.
Thirdly, model performance must be continuously audited to detect drift and unintended correlations (e.g., when new groups are added).
Another important caveat is to calibrate the thresholds for alert generation conservatively in order to reduce false positives and prevent any unnecessary escalations.

\descr{Limitations.}
We also foresee several limitations that warrant consideration with our work.
First, the reliance on Telegram messages introduces inherent biases stemming from platform specific user populations, language variation, and reporting of cyber incidents.
Not all attack types or threat actors are equally represented in Telegram channels, which may cause the model to overfit to the communication patterns of highly active groups rather than the broader threat landscape.
The semantic profiles generated from daily text streams are also sensitive to noise and informal language which may distort the true signal.
Another limitation lies in the alignment between Telegram derived features and real-world incident labels.
Cyber incidents often have delayed, incomplete, or ambiguous reporting timelines, complicating the temporal matching between online chatter and ground truth events.
Additionally, while GraphSAGE captures relational and temporal patterns, the graph itself is constructed with limited knowledge of verified operational chains, meaning that certain inferred relationships may be correlational rather than causative.
Finally, the integration of LLM embeddings introduces computational overhead and dependency on pretrained language models, which may not generalize well across languages or domains without fine-tuning.

\descr{Future Works.}
We envision future works to expand our method in several meaningful directions.
One promising direction involves constructing multi-platform threat graphs by incorporating data from additional sources such as dark web forums, Discord, Twitter (or X), GitHub security advisories, and vulnerability disclosure feeds.
This would allow the model to generalize across diverse communication ecosystems and reduce platform bias.
Another extension involves enhancing temporal modeling through event sequence learning, such as transformers or temporal graph networks, which could better capture the lag structure between early chatter, exploit development, and real-world incidents.
Other future directions could be to fine-tune or domain-adapt LLMs specifically, enabling more accurate interpretation of slang, codewords, and multilingual content prevalent in underground discussions.

\section{Conclusion}
In this work, we introduce \name, a hybrid predictive framework that integrates large language model embeddings with graph neural network-derived temporal and relational features.
We construct daily semantic profiles from 365k Telgram messages and use a combination of GraphSAGE relational embeddings with text-based embeddings to align Telegram data with real-world cyberincidents.
\name demonstrates an F1-score of 0.89 underscoring the importance of language level meaning, temporal dependencies and networked interactions in the cyber threat landscape.
As cyber adversaries continue to evolve in speed and sophistication, the integration of LLMs and GNNs offers a promising path for developing early-warning systems that strengthen the digital infrastructures.

\descr{Acknowledgements.}
This work was supported in part by National Science Foundation grant 212720.

\small
\bibliographystyle{apalike}
\bibliography{refs}

\end{document}